\documentclass{article_saj}
\usepackage{graphicx,saj,multicol,subeqnarray}
\usepackage{multirow}
\usepackage{graphicx}
\usepackage{natbib}
\usepackage{float}
\usepackage{xcolor}
\usepackage{widetext}
\usepackage{url}
\usepackage{bm}
\usepackage{tikz} 
\usepackage{pifont} 
\usepackage{amsfonts}
\usepackage{amssymb}
\usepackage{amsmath,upgreek}
\usepackage{titlesec}
\titlelabel{\thetitle.\quad}
\definecolor{xlinkcolor}{cmyk}{1,0.6,0,0}
\usepackage[bookmarks=false,         
     pdfnewwindow=true,      
     colorlinks=true,    
     linkcolor=xlinkcolor,     
     citecolor=xlinkcolor,     
     filecolor=xlinkcolor,  
     urlcolor=xlinkcolor,      
final=true
]{hyperref}

\def\papertype{\ \hfill\ Preliminary report}

\def\udc{52}
\renewcommand{\thefootnote}{\fnsymbol{footnote}}

\begin{document}
\parindent=.5cm
\baselineskip=3.8truemm
\columnsep=.5truecm
\newenvironment{lefteqnarray}{\arraycolsep=0pt\begin{eqnarray}}
{\end{eqnarray}\protect\aftergroup\ignorespaces}
\newenvironment{lefteqnarray*}{\arraycolsep=0pt\begin{eqnarray*}}
{\end{eqnarray*}\protect\aftergroup\ignorespaces}
\newenvironment{leftsubeqnarray}{\arraycolsep=0pt\begin{subeqnarray}}
{\end{subeqnarray}\protect\aftergroup\ignorespaces}
%


\markboth{\eightrm Galaxy Morphology Classification: Are Stellar Circularities Enough?} 
{\eightrm \MakeUppercase{K. Baucalo} and \MakeUppercase{A. Mitra{\v s}inovi{\' c}}}

\begin{strip}

{\ }

\vskip-1cm

\publ

\type

{\ }


\title{\MakeUppercase{Galaxy Morphology Classification: Are Stellar Circularities Enough?}}


\authors{Katarina Baucalo$^{1}$ and Ana Mitra{\v s}inovi{\' c}$^{2}$}

\vskip3mm


\address{$^1$Department of Astronomy, Faculty of Mathematics,
University of Belgrade\break Studentski trg 16, 11000 Belgrade,
Serbia}


\Email{kataarina981@gmail.com}

\address{$^2$Astronomical Observatory, Volgina 7, 11060 Belgrade, Serbia}

\Email{amitrasinovic@aob.rs}


\dates{May 13, 2025}{June 23, 2025}


\summary{We present a preliminary study exploring whether the stellar orbital circularity of simulated galaxies, available from precomputed catalogs in the IllustrisTNG project, can be used as a proxy for broad morphological classification. We focus on the publicly available "Stellar Circularities, Angular Momenta, Axis Ratios" catalog, which enables a simple kinematic decomposition of the stellar component into disk and spheroid subsystems. By validating this approach against the more detailed five-component kinematic decomposition in TNG50, we confirm that the circularity-based disk fraction correlates most strongly with the thin disk, while the bulge fraction broadly represents the combined contribution of classical bulges and stellar halos. We then apply this decomposition to galaxies in the TNG100 simulation at redshift $z=0$ and identify a data-motivated threshold of $\mathrm{F_{disk}} = 0.25$ to distinguish early- and late-type galaxies. This threshold, lower than the commonly adopted value of $0.4$, better captures the diversity of disk-dominated systems and avoids excluding galaxies with moderately prominent disks. Additionally, we identify irregular or morphologically complex systems based on galaxies with low total disk and spheroid mass fractions. Using this classification, we recover a morphology–density relation that is broadly consistent with observations: late-type galaxies dominate in the field, while early-type galaxies are the most prevalent morphological type in clusters. Our results demonstrate that stellar circularity alone can serve as an accessible and computationally efficient morphological proxy. We also discuss the potential for this classification to support machine learning efforts as a baseline or training set for future morphological studies. }



\end{strip}

\tenrm

\renewcommand{\thefootnote}{\arabic{footnote}}


\section{INTRODUCTION}

\indent Galaxies are gravitationally bound systems composed of stars, interstellar gas, and dust, embedded within a much larger invisible structure known as a dark matter halo \citep[e.g.,][]{mo&vdBosch&White2010}. Each galaxy is a unique system that varies in stellar mass, gas content, and overall structure. The earliest classifications of galaxies emerged from direct observations, primarily grounded in morphological differences such as shape and internal structure. Although Edwin Hubble introduced the first widely adopted morphological classification scheme \citep{Hubble1926ApJ}, shortly after it was established that galaxies are extragalactic systems distinct from the Milky Way, his system is not exhaustive. Several alternative classification frameworks have since been proposed \citep[e.g.,][]{deVac1959morf, morgan1957, morgan1969}, yet the Hubble sequence remains the most widely used and recognized in contemporary astronomy. Despite its various subtypes, the Hubble classification broadly divides galaxies into three main morphological types: elliptical, spiral, and irregular galaxies.\\
\indent The study of galaxy morphology extends far beyond the pursuit of a systematic classification scheme. Different morphological types encode crucial information about the formation history of a galaxy, and analyzing morphology can also yield insights into its future evolutionary path. Just as each individual star follows a characteristic evolutionary trajectory, so too do the structural components of galaxies (such as disks, bulges, and bars) evolve over time. These components often undergo co-evolution, interacting dynamically and chemically in ways that shape the global properties of a galaxy. Moreover, there exists a strong correlation between the morphological type of a galaxy and its physical characteristics, including color, gas content, and star formation rate, as well as the environment in which it resides \citep[e.g.,][]{dressler1980tsigmadata, dressler1980tsigma, Blanton+Moustakas2009}. For instance, late-type galaxies tend to be gas-rich and actively star-forming, whereas early-type galaxies are typically red, gas-poor, and quiescent. Environmental influences, ranging from isolated field conditions to dense cluster environments, are critical in shaping these trends\footnote{The connection between the local environment and morphology of a galaxy is known as a morphology-density relation.}. Consequently, studying the galaxy morphology is central to answering fundamental questions about the formation and evolution of galaxies \citep{Conselice2014}.\\
\indent Modern cosmological hydrodynamical simulations provide a powerful framework for testing theoretical predictions and bridging the gap between theory and observations. In the context of cosmological simulations, it is crucial to accurately reproduce observed phenomena. So far, simulations have proven to be quite useful in advancing our understanding of various topics and trends related to the evolution of galaxies and their interactions, properties, and scaling relations \citep[e.g.,][]{Vogelsberger+2020NatRP...2...42V, Crain+vaddeVoort2023ARA&A..61..473C}.  In this work, we will utilize the IllustrisTNG cosmological simulations\footnote{\url{https://www.tng-project.org/data/}} \citep{TNGmethods2017, TNGmethods2018, Nelson+2019ComAC}.\\
\indent When it comes to morphology, despite the richness of the data, morphological classification in simulations remains methodologically challenging. Prior efforts have primarily relied on photometric proxies through mock observations \citep[e.g.,][]{Rodriguez-Gomez+2019, Huertas-Company+2019, Varma+2022, Gong+2025} or complex kinematic decomposition techniques, such as Gaussian mixture models applied to stellar orbits \citep{Du+2019, Du+2020}. While these approaches offer valuable precision, they often require extensive computation or post-processing, which can limit scalability and typically represent a research study on their own. Although some of these studies have released publicly available catalogs, the datasets are often limited to one or a few specific snapshots of the cosmological simulation, focused on specific IllustrisTNG simulation boxes, or encompass a fairly conservative sample of galaxies. Hence, studies that aim to investigate specific morphological types of galaxies in more detail, or the dependence of specific global parameters (e.g., gaseous content, star formation rates, environmental proxies) on morphological types, have to either perform morphological classifications of galaxies from scratch or rely on available supplementary catalogs. The smallest IllustrisTNG simulation box, TNG50 \citep{TNG50-1-2019, TNG50-2-2019}, includes a supplementary catalog containing data on the fractional contributions of various subsystems of the stellar component of galaxies. The name of this catalog is (t) \emph{Galaxy Morphologies (Kinematic) and Bar Properties} \citep{Zana+2022}, hereafter referred to as catalog (t). The fractional contributions of components were derived using the complex kinematic decomposition tool \textsc{Mordor}\footnote{\url{https://github.com/thanatom/mordor}} for galaxies that have sufficient resolution (at least 1000 stellar particles) at any given snapshot of the simulation.\\
\indent In contrast to the observed galaxies, the morphology of the simulated galaxies can be determined easily via kinematic decomposition, since simulated galaxies are composed of particles, and the data include all the necessary parameters for such a procedure. The simplest kinematic decomposition is based on determining its binding energy $E$ and the circularity of its orbit $\epsilon$ for each stellar particle in the kinematic phase space. For each stellar particle, the orbital circularity parameter $\epsilon$ is defined as the ratio $j_z/j_c(E)$, where $j_z$ is the component of its angular momentum along the $z$-axis (assuming the angular momentum vector of the galaxy is aligned with the positive direction of the $z$-axis), and $j_c(E)$ is the maximum angular momentum a particle can have at a given energy $E$, assuming axial symmetry of the gravitational potential in the galactic plane\footnote{In other words, $j_c(E)$ corresponds to the angular momentum of a particle with the same energy $E$ that moves on a perfectly circular orbit.}. The value of $\epsilon$ ranges from $-1$ to $1$, depending on whether the rotation of the particle is opposite to (negative values) or aligned with (positive values) the overall direction of rotation in the system. The energy $E$ represents the total energy of the stellar particle and serves as a measure of its gravitational binding to the system (assuming the total energy is negative).\\
\indent The \textsc{Mordor} software for the kinematic decomposition enables division of the stellar component into five subsystems based on complex criteria for $\epsilon$ and $E$, which was performed in the catalog (t) exclusively within the TNG50 simulation box, as mentioned previously. Although the software can be used for larger simulation boxes, this task can be resource-intensive and time-consuming. A simpler kinematic decomposition allows splitting the stellar component into only two subsystems (broadly defined and referred to as a bulge and a disk), which is even more efficient for larger samples\footnote{This is because it does not require galaxies to have a very high particle count; decomposition can be performed even on galaxies consisting of just a few hundred particles.}. This simpler method was applied in the supplementary catalog (c), named \emph{Stellar Circularities, Angular Momenta, Axis Ratios} \citep{Genel+2015}, hereafter referred to as catalog (c). This catalog is available for all simulation boxes and every simulation snapshot. For this reason, catalog (c) provides highly accessible information that should allow at least broad morphological classification, but we argue that it is unfortunately very underutilized.\\
\indent In this preliminary study, we leverage this catalog (c) to investigate whether stellar disk fractions, as defined by circularity-based decomposition, can serve as a quantitative morphological proxy for classifying galaxies into early- and late-type systems. As an exploratory extension and a certain test, we also examine whether this simple classification scheme is sufficient to recover the morphology-density relation in the simulated universe. Our goal is twofold: to assess the viability of circularity-based morphology classification as a lightweight alternative to more complex methods and to test whether IllustrisTNG reproduces a key observational trend (i.e., at least a broad agreement with the morphology-density relation) without requiring mock image synthesis, detailed orbit modeling, or any other complex method. By focusing on simplicity, we aim to lay the groundwork for more detailed follow-up studies and to make morphology analyses more accessible.\\
\indent This paper is organized as follows. In Section~\ref{methods}, we describe the simulation boxes that we use and explore the way in which we can utilize the parameters available in the catalog (c) most efficiently. In Section~\ref{results}, we attempt to find a data-informed and physically meaningful criterion for the broad morphological classification of galaxies, and test the classification scheme while exploring the environmental dependence of different morphological types. In Section~\ref{discussion}, we discuss our results in the broader context of other similar and relevant studies. Finally, we give the concluding remarks in Section~\ref{conclusion}. 

\section{METHODS}\label{methods}

\indent The IllustrisTNG cosmological hydrodynamical simulations of galaxy formation were carried out using the \texttt{Arepo} code \citep{Springel2010AREPO} and adopt the cosmological parameters from \citet{PlanckColab+2016}: a matter density of $\Omega_\mathrm{m} = 0.3089$, baryon density $\Omega_\mathrm{b} = 0.0486$, dark energy density $\Omega_\Lambda = 0.6911$, Hubble constant $H_0 = 67.74\; \mathrm{km}\; \mathrm{s}^{-1}\; \mathrm{Mpc}^{-1}$, power spectrum normalization $\sigma_8 = 0.8159$, and primordial spectral index $n_\mathrm{s} = 0.9667$. These simulations started at redshift $z = 127$, and the results are stored in 100 snapshots ranging from $z = 20$ to $z = 0$. They incorporate key astrophysical processes, including gas cooling, feedback from supernovae and active galactic nuclei (AGN), and large-scale structure formation, allowing galaxies to evolve self-consistently within cosmological volumes.\\
\indent The simulation suite includes three flagship runs (TNG50, TNG100, and TNG300) named approximately after the side length of the simulation box (i.e., 50, 100, and 300 Mpc, respectively). In addition to differences in volume, the three boxes also differ in particle resolution, enabling studies at various scales and levels of detail. The IllustrisTNG team highlights that the TNG100 \citep{Marinacci+2018, Naiman+2018, Nelson+2018, Pillepich+2018, Springel+2018} offers an optimal balance between volume and resolution. The mass resolution for the TNG100 is $7.5 \times 10^{6}\;\mathrm{M}\odot$ for dark matter particles and approximately $1.4 \times 10^{6}\;\mathrm{M}\odot$ for baryonic particles \citep{Springel&Hernquist+2003,TNGmethods2018}. We will use this simulation box, since it is often viewed as the main box, and considering that it features a diversity of environments (i.e., multiple galaxy clusters and a higher number of galaxy groups) in contrast to the smaller simulation box with higher resolution (the TNG50), which contains only one Virgo-like cluster and a small number of galaxy groups. Throughout this paper, we will focus solely on the last, present-day snapshot, that is, the redshift $z=0$.\\
\indent Before we dive deeper into analysis, we will first compare two supplementary catalogs, catalog (c) and catalog (t), in the TNG50 simulation box to understand and validate the physical meaning behind the parameters from catalog (c) that we will use throughout this study for morphological classification.

\subsection{Validation of mass fractions of different structures from supplementary catalogs}

\begin{figure}[!h]
	\centerline{\includegraphics[width=0.99\columnwidth, keepaspectratio]{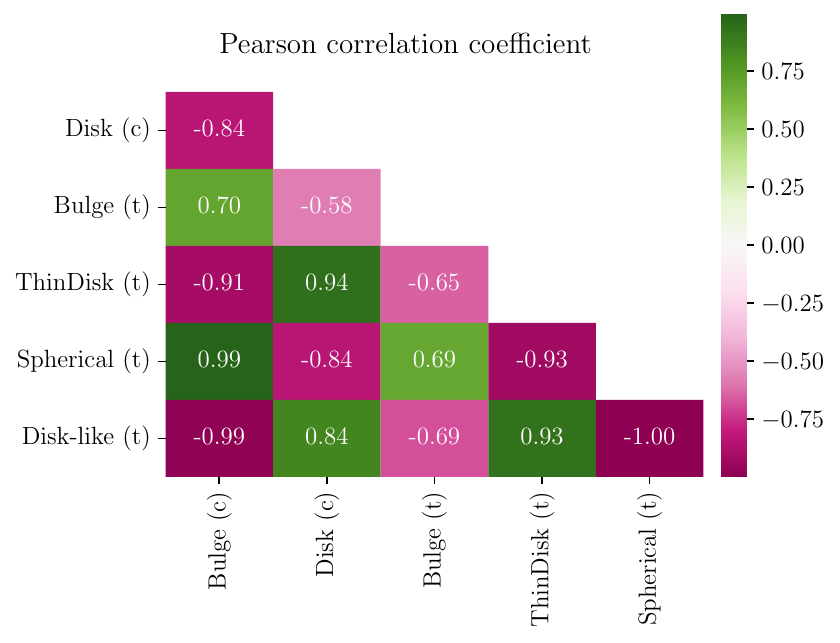}}
	\caption{Correlation between fractions of different structures (or their sum) in supplementary catalogs (c) and (t), represented through Pearson correlation coefficient. The exact values of the correlation coefficient are given in each cell.}
	\label{fig-correlation}
\end{figure}

\indent Catalog (c) contains information on mass fractions of only two stellar subsystems, referred to as a bulge and a disk. A bulge is defined as the cumulative mass of all particles with $\epsilon<0$ multiplied by two, which is noted as a common way to define a bulge. The name of this parameter in the catalog is \texttt{CircTwiceBelow0Frac}. However, as \citet{Zana+2022} rightfully pointed out, this definition often does not differentiate between a bulge and a stellar halo and, without additional examination, should represent the mass fraction of both spherical components. For the disk component, the catalog contains two mass fraction parameters. In a simpler way, a disk is defined as a cumulative mass of all particles with $\epsilon>0.7$. The second parameter that represents the mass fraction of the stellar disk is corrected by subtracting the contribution of a bulge, under the assumption that the distribution is symmetric around $\epsilon = 0$. Specifically, it is defined as the mass fraction of stellar particles with $\epsilon > 0.7$, reduced by the mass fraction of those with $\epsilon < -0.7$. The name of this second parameter is \texttt{CircAbove07MinusBelow07Frac}, and we adopt this in this study as F$_\mathrm{disk}$. Based on its definition, it should represent a thin disk structure more accurately, not all disk-like components, as they can contain particles with lower circularities. These parameters are calculated twice; one approach takes into account stellar particles inside 10 times the stellar half-mass radius, while the other takes into account all stellar particles in the subhalo. Throughout this work, we use the former. This range encompasses the majority of stellar particles, although some potential tidal features are excluded.\\
\indent In contrast, catalog (t) contains information on mass fractions of as many as five different components: thin and thick disk, pseudo-bulge, classical bulge, and stellar halo. This was made possible by the elaborate analysis included in the \textsc{Mordor} software and described in detail in \citet{Zana+2022}. The decomposition has proven remarkably successful and precise, and we will use it to compare two catalogs. We are primarily interested in validating our initial assumptions that the bulge fractions in a catalog (c) correspond to both spherical components (i.e., bulge and stellar halo) and that the disk fraction is more in line with only a thin disk. Perhaps counter-intuitively and despite the name, pseudo-bulge is one of the disk-like components, as pseudo-bulges form from the disk itself during a secular evolution \citep[e.g.,][]{falcon2012secular-book, sellwood2014RvMP} and are kinematically similar to disks \citep[see, for example, the review by][]{Kormendy-Kennicutt2004}.\\
\indent We start by calculating Pearson correlation coefficients between the two parameters from the catalog (c) and the relevant parameters or their sum from the catalog (t). The results are shown in Fig.~\ref{fig-correlation}. The disk-like (t) parameter represents the sum of all disk-like structures in catalog (t): thin and thick disk, and pseudo-bulge, while the spherical (t) parameter represents the sum of the classical bulge and stellar halo.

\begin{figure}[!h]
	\centerline{\includegraphics[width=0.99\columnwidth, keepaspectratio]{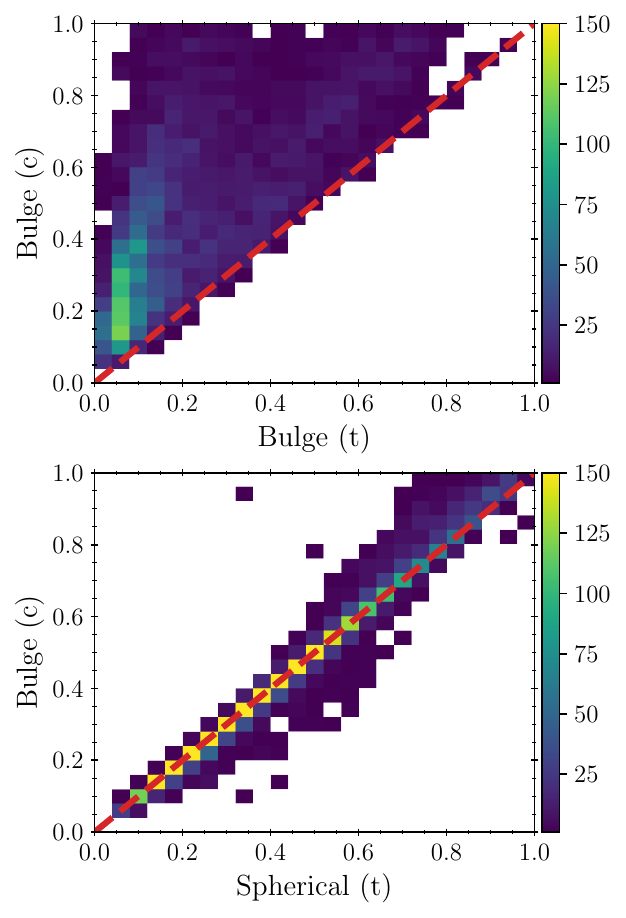}}
	\caption{Correlations between bulge fraction in catalog (c) and bulge/spherical fractions in catalog (t): upper/lower panel. The red dashed line represents the $y=x$ line.}
	\label{fig-corr-bulge}
\end{figure}

\indent It can be unambiguously concluded that the parameter labeled "bulge" in catalog (c) corresponds to the total mass of all spherical substructures, not just the classical bulge. This is because an analysis based solely on orbital circularities cannot distinguish between bulge and halo components, as both are characterized by nearly radial orbits, i.e., motion dominated by velocity dispersion rather than rotation. For the disk component, the distinction is less obvious. However, it can still be concluded that the parameter in catalog (c) shows a stronger correlation with the thin disk fraction than with the combined mass of all disk-like components.\\
\indent Since the Pearson correlation coefficient determines the degree to which a relationship between two variables is linear, and not the actual equality between variables, we additionally examine these relationships visually. In Fig.~\ref{fig-corr-bulge} we show two-dimensional distributions of galaxies in the "Bulge (t)" -- "Bulge (c)" plane (upper panel) and in the "Spherical (t)" -- "Bulge (c)" plane (lower panel). The lower panel shows a clear linear relationship between the bulge component in catalog (c) and the sum of all spherical structures in catalog (t). In contrast, the upper panel of Fig.~\ref{fig-corr-bulge} does not reveal such a linear correlation between the bulge parameters in the two catalogs, since the catalog (c) values are significantly higher. This discrepancy arises because the bulge in catalog (c) effectively represents the total contribution from all spherical structures, as demonstrated in the preceding analysis. The scatter in the lower panel is generally low, although at least one galaxy appears as an extreme outlier where the spherical fraction is significantly overestimated. This extreme outlier has a significant portion of its stellar mass, more than $40\%$, assigned to the pseudo-bulge, which is perhaps why its spherical fraction in the catalog (c) is overestimated. Other, less extreme, outliers also have somewhat prominent pseudo-bulges and typically have higher thick than thin disk fractions.

\begin{figure}[!h]
	\centerline{\includegraphics[width=0.99\columnwidth, keepaspectratio]{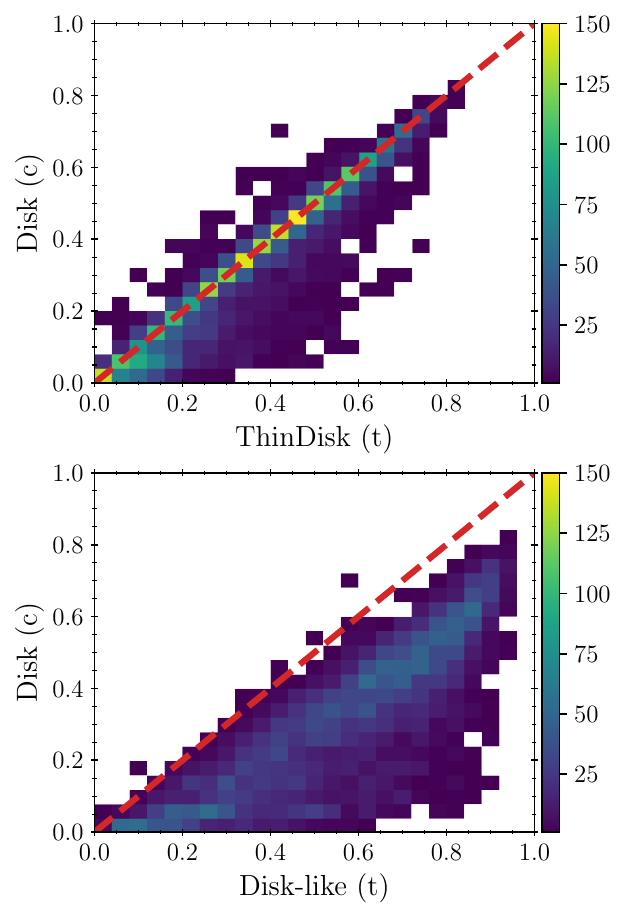}}
	\caption{Similarly to Fig.~\ref{fig-corr-bulge}, correlations between disk fraction in catalog (c) and thin disk/disk-like structures fractions in catalog (t): upper/lower panel. The red dashed line represents the $y=x$ line.}
	\label{fig-corr-disk}
\end{figure}

\indent Similarly, in Fig.~\ref{fig-corr-disk} we show two-dimensional distributions of galaxies in the "ThinDisk (t)" -- "Disk (c)" plane (upper panel) and in the "Disk-like (t)" -- "Disk (c)" plane (lower panel). In the upper panel, when examining the correlation between the disk component from the catalog (c) and the thin disk from the catalog (t), a clear linear trend is observed with mild scatter. In contrast, the correlation between the disk in catalog (c) and the sum of all disk-like structures from catalog (t) shows only a weak linear dependence accompanied by significant scatter. Notably, the disk fraction in catalog (c) is almost always lower than the total fraction of disk-like structures in catalog (t). Although the disk in catalog (c) is a good proxy for the thin disk component, a scatter in the upper panel of Fig.~\ref{fig-corr-disk} suggests that the relationship is not as tight as it is with bulge and spherical components. However, the scatter does not appear symmetrical, and it appears that the disk in catalog (c), when imprecise, more often underestimates the thin disk. We will consider this when we discuss the possible cut-off values of $\mathrm{F}_\mathrm{disk}$ for the morphological classification.

The main conclusion of this analysis is that it enables a clearer understanding of the parameters within the catalog (c), providing a solid foundation for the subsequent analysis to be conducted within the TNG100 simulation box. With the physical meaning of the catalog (c) parameters now established, we can approach the primary objective of this project, the morphological classification of galaxies based on the known mass fractions of different components obtained through a simple precomputed kinematic decomposition.

\section{RESULTS}\label{results}

\begin{figure}[!h]
	\centerline{\includegraphics[width=0.99\columnwidth, keepaspectratio]{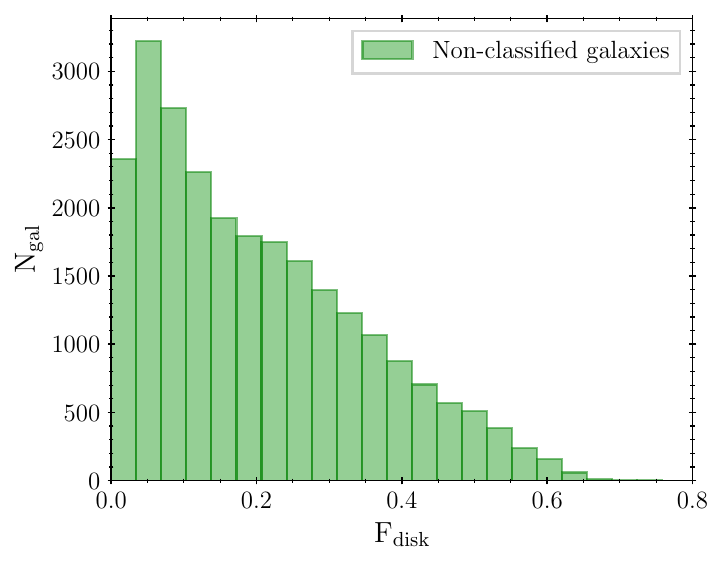}}
	\caption{Distribution of disk fraction for unclassified early-type or late-type galaxies. Irregular galaxies, which were previously classified, are not included in this figure.}
	\label{fig-nonclassified}
\end{figure}

\indent Having justified the use of the catalog (c) and established the physical interpretation of the parameters \texttt{CircAbove07MinusBelowNeg07Frac} and \texttt{CircTwiceBelow0Frac}, we subsequently used these quantities as indicators of thin disk ($\mathrm{F}_\mathrm{disk}$) and spherical structures ($\mathrm{F}_\mathrm{sph}$), respectively. In principle, the combined fraction of spherical and disk-like components should not exceed unity, since by definition, the fractional contribution of any component must not exceed the whole. However, there are instances where this condition is not met. In such cases, the spherical components are typically dominant and very likely overestimated; these systems are almost certainly spheroidal or elliptical galaxies.\\
\indent Conversely, there are also cases where the sum of the two components falls significantly below one. This can be explained in two ways. First, a substantial fraction of the stellar mass may reside in substructures that are not gravitationally bound to the galaxy. Second, and perhaps a more likely explanation, a considerable number of stellar particles may be on highly eccentric orbits, which do not belong to well-defined spherical components and, due to their eccentricity, are also excluded from the disk category. For this reason, it is justified to treat galaxies where the combined fraction of disk and spherical components is significantly below one (in this study, we use a threshold of $2/3$) as irregular or morphologically complex systems. When selecting this threshold, it is important to recognize that $\mathrm{F}_\mathrm{disk}$ does not fully account for all disk-like structures and that even thin disks can often be underestimated. Therefore, adopting a slightly lower threshold is more conservative, helping to exclude galaxies with a prominent thick disk or pseudo-bulge. Additionally, decreasing this threshold would contaminate our sample of galaxies (to be classified into early- or late-type) with many genuinely irregular galaxies.\\
\indent At the very beginning, we have classified a part of the sample into "irregular/complex" systems, as described above. The remaining unclassified galaxy sample may include elliptical, disk-like, and even lenticular (S0) galaxies. In an ideal scenario, where this sample is composed of well-separated subpopulations of ellipticals and disk galaxies, the one-dimensional distribution of $\mathrm{F}_\mathrm{disk}$ (or, equivalently, $\mathrm{F}_\mathrm{sph}$, though this study focuses on the disk fraction) would be expected to exhibit a bimodal shape. This expectation stems from the assumption that elliptical and disk galaxies are more numerous than lenticular ones, representing a transitional morphological type. However, as shown in Fig.~\ref{fig-nonclassified}, the resulting one-dimensional distribution is not bimodal, but rather monotonically decreasing.

\begin{figure}[!h]
	\centerline{\includegraphics[width=0.99\columnwidth, keepaspectratio]{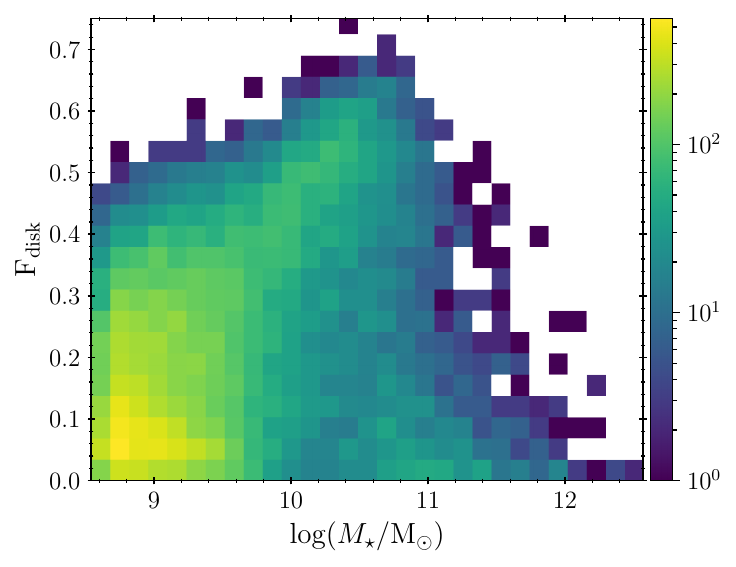}}
	\caption{Diagram of disk fraction and logarithmic stellar mass, where the number of galaxies is also logarithmic (presented on the colorbar).}
	\label{fig-logM-fdisk}
\end{figure}

\indent The deviation from the ideal case is most likely due to the large number of low-stellar-mass galaxies (dwarf galaxies), which are predominantly elliptical or spheroidal. As a result, the distribution is not bimodal, making it unsuitable for determining an optimal threshold to separate early- and late-type morphological systems using this approach. This interpretation is supported by Fig.~\ref{fig-logM-fdisk}, which shows the two-dimensional distribution of the stellar mass versus the disk fraction, plotted as $\log(M_\star/\mathrm{M}_\odot)$ against $\mathrm{F_{disk}}$. The plot reveals that galaxies with high disk fractions predominantly occupy the stellar mass range $10 \leq \log(M_\star/\mathrm{M}_\odot) \leq 11$, while a large number of early-type galaxies, with very low disk fractions, are also clearly visible, especially in the low-mass regime.

\begin{figure}[!h]
	\centerline{\includegraphics[width=0.99\columnwidth, keepaspectratio]{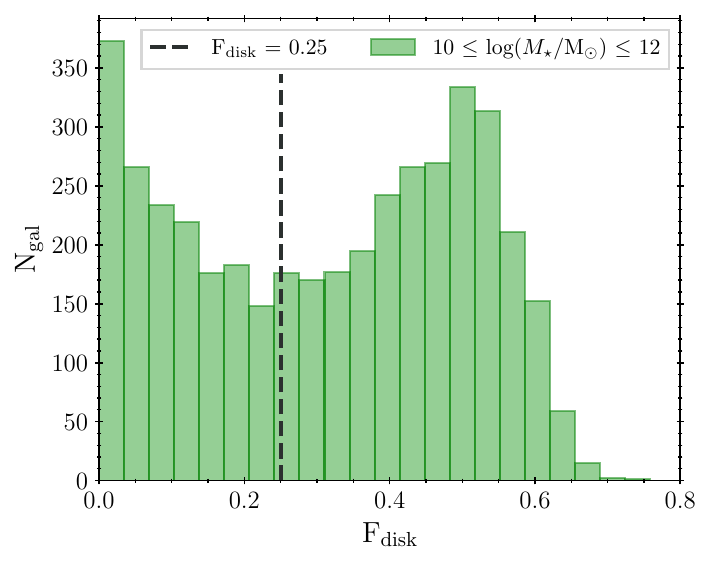}}
	\caption{Disc fraction distribution for unclassified galaxies in the range $10 \leq \mathrm{log}(M_\star / \mathrm{M}_\odot) \leq 12$. The black dashed line represents the cut-off value for the disk fraction, where galaxies with a disk fraction higher than the cut-off are late-type, and vice versa for early-type.}
	\label{fig-fdisk-localmin}
\end{figure}

\indent Since one approach to defining a boundary between early- and late-type galaxies involves identifying a sample whose one-dimensional $\mathrm{F_{disk}}$ distribution is bimodal, it is clear that the stellar mass range $10 \leq \log(M_\star/\mathrm{M}_\odot) \leq 12$ holds potential for such a distribution. Extending this interval toward lower masses would introduce a large number of dwarf galaxies into the sample, systems in which $\mathrm{F_{disk}}$ is typically low and the spheroidal component is dominant. In fact, this expectation is supported by the data shown in Fig.~\ref{fig-fdisk-localmin}.

\begin{figure}[!h]
	\centerline{\includegraphics[width=0.99\columnwidth, keepaspectratio]{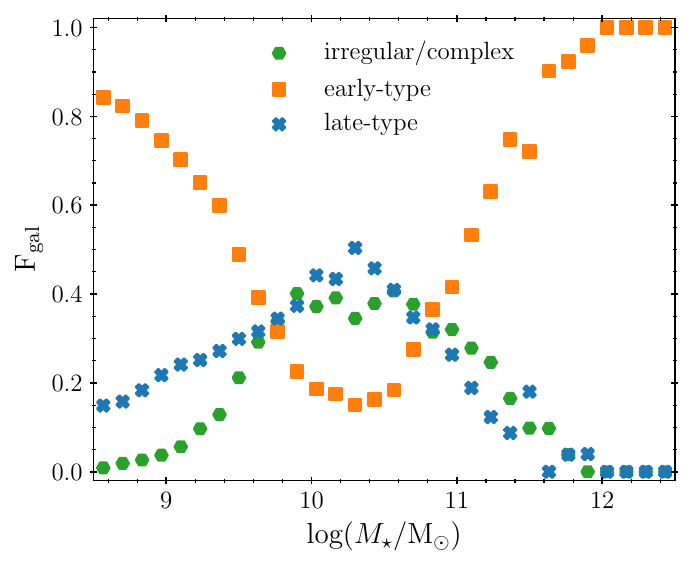}}
	\caption{The fraction of galaxies per mass bin as a function of the logarithm of stellar mass. Galaxy types are color-coded: green for irregular/complex, orange for early-type, and blue for late-type galaxies (as indicated by the legend).}
	\label{fig-logM-fgal}
\end{figure}

\indent In Fig.~\ref{fig-fdisk-localmin}, a clear bimodal distribution is visible, along with a local minimum located at a disk fraction of approximately $0.225$. Based on this, we chose a slightly higher threshold of $\mathrm{F_{disk}} = 0.25$ since it represents a reasonable value, at approximately a similar distance from the two peaks. Accordingly, all galaxies with a disk fraction greater than $0.25$ are identified as late-type systems, while the remaining unclassified galaxies are considered to be of an early morphological type. It is also worth noting that Fig.~\ref{fig-fdisk-localmin} allows a tentative classification of lenticular galaxies (S0). These galaxies would be characterized by $\mathrm{F_{disk}}$ in a certain range centered around our adopted threshold. However, such a classification was not pursued here, as this level of analysis falls outside the scope of the present study. Instead, we recognize only early- and late-type galaxies, and lenticular galaxies, given that they represent a transitional morphological type, can fall into both categories, depending on how prominent their disks are.\\
\indent In Fig.~\ref{fig-logM-fgal}, we show the fraction of galaxies of each morphological type (irregular/complex, early-type, and late-type) as a function of stellar mass per mass bin, based on the classification scheme developed in this study. Each stellar mass bin contains galaxies normalized to unity, allowing us to compare relative abundances within bins. The plot reveals that irregular or morphologically complex galaxies (green) have a fairly symmetric distribution and occupy the intermediate mass regime, indicating a broad range of dynamical states and formation pathways. Late-type galaxies (blue) are most common in the intermediate-mass range and have a unimodal distribution that is skewed and asymmetric, featuring more low-mass galaxies than massive ones. In contrast, early-type galaxies (orange) dominate both the high-mass and low-mass ends of the distribution, having a bimodal distribution. Moreover, they are not just dominant, but perhaps the exclusive morphological type at the high-mass end. This is in agreement with observational knowledge since the most massive known galaxies are core Dominant (cD) galaxies \citep[e.g.,][]{morgan1957, morgan1969} located at the centers of rich galaxy clusters \citep[e.g.,][]{Matthews1964}.

\indent The trends we clearly see in Fig.~\ref{fig-logM-fgal} reflect how morphology correlates with stellar mass in the TNG100 simulation and support the rationale for selecting the intermediate-mass regime when identifying a bimodal distribution in disk fraction.

\subsection{Environmental considerations}

\begin{figure}[!h]
	\centerline{\includegraphics[width=0.99\columnwidth, keepaspectratio]{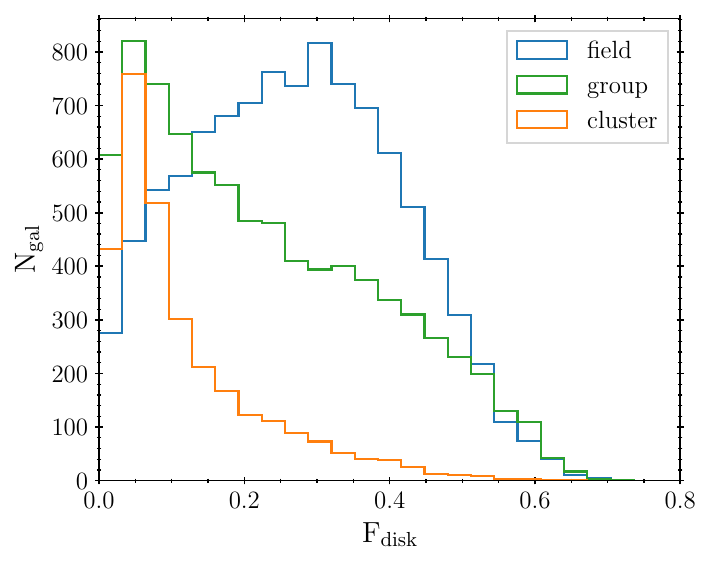}}
	\caption{Distribution of disk fraction of galaxies for different environments. Due to the contamination of the total sample with a large number of early-type galaxies, a subsample in the mass interval $9 \leq \log(M_\star/\mathrm{M_\odot}) \leq 12$ was considered. Environment types are color-coded: blue for field, green for group, and orange for cluster (as indicated by the legend).}
	\label{fig-fdisk-env}
\end{figure}

\indent One way to validate the galaxy classification, and one of the goals of our study, is to examine the type of environment in which different morphological types are found. Environmental data are provided in the Group catalog, based on properties of identified "Friends-of-Friends" halos \citep[that is, host halos, see e.g.,][]{More+2011}, and the division into clusters, groups, and field galaxies follows the criteria outlined in \citet{Paul+2017}. A system is classified as a cluster if the mass of its host halo exceeds $8 \times 10^{13}\;\mathrm{M_{\odot}}$. Note that this applies to our study, which is focused on redshift $z=0$, as it is not applicable at high redshifts, where proto-clusters of lower mass may exist. Field galaxies are defined as those that reside in halos with a mass below $5 \times 10^{12}\;\mathrm{M_{\odot}}$ and hosting fewer than $30$ subhalos, ensuring that the system is truly isolated. This additional criterion for field galaxies allows for the existence of small groups with somewhat lower halo masses but a relatively high number of subhalos, such as associations of dwarf galaxies or low-mass groups with only one fairly massive galaxy and a high number of satellites. Any galaxy not meeting the criteria for either clusters or fields is classified as residing in a group environment. This procedure constitutes a relatively coarse but practical environmental classification, providing a simple framework for investigating environmental trends. Due to contamination of the total sample with a large number of early-type galaxies at both mass ends (as we saw in Fig.~\ref{fig-logM-fgal}), for environmental consideration, we focus on a subsample in the mass interval $9 \leq \log(M_\star/\mathrm{M_\odot}) \leq 12$. \\
\indent In accordance with this framework, we analyzed the disk fraction $\mathrm{F_{disk}}$ as a function of the type of environment, as shown in Fig.~\ref{fig-fdisk-env}. It is evident that the disk fraction $\mathrm{F_{disk}}$ shifts toward higher values in lower-density environments. 

\indent The distribution of galaxies of different morphological types across various environments was also examined. The results of this analysis are presented in Table~\ref{tab1}.\\

\begin{table}[!h]
\centering
\caption{Number of galaxies per morphological type for different types of environments. The total number of galaxies per environment type is also shown in boldface font.}
\label{tab1}

\begin{tabular}{c|c|c|c}
                   & cluster & field & group \\ \hline
early-type         & 1481    & 4233  & 3553  \\ \hline
irregular/complex  & 1381    & 1188  & 2440  \\ \hline
late-type          & 121     & 4500  & 2140  \\ \hline 
\textbf{total} & \textbf{2983}    & \textbf{9921}  & \textbf{8133} \\ 
\end{tabular}%

\end{table}

\indent Based on the data in Table~\ref{tab1}, which shows the number of galaxies per morphological type and environment, it is fairly evident that early-type galaxies are still the most numerous. This can be explained by the fact that the redshift of the sample is $z=0$, which corresponds to the present-day Universe. Early-type galaxies, particularly ellipticals, are expected to dominate at such low redshifts, as is reflected in the results. On the other hand, the distribution of morphological types across the three different environment categories is consistent with the previously discussed morphology–density relation. Specifically, the fraction of late-type galaxies decreases accordingly with increasing environmental density. A particularly notable trend is observed in cluster environments, where early-type and irregular/complex galaxies dominate, and late-type galaxies are nearly absent. Irregular/complex galaxies are most prevalent in groups, probably because of the more diverse and frequent interactions with other galaxies. Additionally, irregular/complex galaxies are slightly less common in the field than in groups or clusters. Even then, field irregular/complex galaxies are typically either currently interacting or have had a recent interaction with another galaxy, based on a quick, preliminary assessment of their history and nearest neighbors from subhalo and group catalogs.\\
 \indent Therefore, based on the combined results presented in Fig.~\ref{fig-fdisk-env} and Table~\ref{tab1}, and the trends that we observe with respect to different morphological types, we conclude that there is at least tentative evidence that the morphology-density relation is reproduced in IllustrisTNG.

\section{DISCUSSION}\label{discussion}

\indent A key outcome of our study is the identification of a data-informed threshold for differentiating between early- and late-type galaxies using a simple kinematic decomposition based on stellar circularities, available in an underutilized supplementary catalog. Although several prior studies have adopted this simple approach to approximate morphological classification, the choice of threshold to separate disk-dominated from spheroid-dominated systems has varied. In particular, a frequently cited work \citep{Joshi+2020} applying the same method used a threshold of $\mathrm{F_{disk}} = 0.4$ to identify late-type galaxies. Although the authors themselves acknowledged that this threshold might be too aggressive, as many galaxies with lower disk fractions still appeared to have disk-like structure, later studies often adopted this value without reassessing its empirical justification \citep[e.g.,][]{Galan-deAnta+2022MNRAS.517.5992G, Lokas2022A&A...662A..53L, Fontirroig+2024}. Interestingly, \citet{Park+2022} opt for a rather unusual method; they define the disk as all particles with circularities higher than $\epsilon>0.5$, which is lower than the commonly used criterion of $\epsilon>0.7$, but impose an aggressive disk fraction threshold for late-type galaxies of $\mathrm{F_{disk}} = 0.5$.

\indent In contrast, our analysis favors a lower threshold of $\mathrm{F_{disk}} = 0.25$, derived directly from the bimodal distribution observed in the intermediate stellar mass range $10 \leq \log(M_\star/\mathrm{M}_\odot) \leq 12$, where both disk-dominated and spheroid-dominated galaxies coexist in appreciable numbers. This choice of threshold is not only data-motivated, but also supported by our validation against the more detailed five-component decomposition available in the TNG50 catalog (t), which confirms that the circularity-based disk fraction from the catalog (c) correlates most strongly with the thin disk component, and the thin disk is even often underestimated. A higher threshold, like 0.4, would risk excluding systems with substantial but not dominant disks, including plausible lenticulars or dynamically heated spirals with pronounced thick disks and possible pseudo-bulges, thereby underestimating the late-type fraction and potentially biasing any environmental or evolutionary interpretation.\\
\indent Beyond the internal consistency of our classification scheme, we also find tentative evidence that the morphology–density relation is reproduced in the TNG100 simulation box. Our findings reveal a clear trend in the distribution of morphological types across environments, with late-type galaxies prevailing in the field, whereas early-type and irregular/complex systems are more common in denser regions. However, we emphasize that this should only be considered tentative evidence for the reproduction of the morphology-density relation. This is mainly due to our simplified approach to characterizing the environment. We adopt a broad, three-tier classification (field, group, cluster) based on host halo mass and subhalo count, rather than using local or continuous environmental density measures typically used in studies of the morphology–density relation \citep[e.g.,][]{dressler1980tsigmadata, dressler1980tsigma, Blanton+Moustakas2009}. Consequently, our results cannot be directly compared to studies that quantify the environment through nearest-neighbor metrics or smoothed density fields. Nevertheless, the fact that such broad trends emerge even with this broad classification suggests that the underlying relationship between morphology and environment is robust and encourages further investigation using more detailed environmental measures.\\
\indent This qualitative agreement with the observed morphology–density relation is encouraging, especially given the simplicity of the method. It suggests that circularity-based morphology, despite its limitations, can capture broad environmental trends and may serve as an efficient proxy for large-scale studies. Moreover, while the morphology-density relation was confirmed in EAGLE simulations \citep{Pfeffer+2023MNRAS.518.5260P}, such an analysis or any similar examination was not previously performed with IllustrisTNG. Importantly, these findings validate the utility of catalog (c) beyond its currently limited application and highlight its potential for statistically robust morphology-oriented studies without requiring computationally expensive decompositions or synthetic observations. However, we caution that our classification scheme remains broad and does not distinguish between finer morphological classes, such as lenticulars and barred spirals, or otherwise morphologically complex systems.\\
\indent Furthermore, given that the catalog (c) is available across all snapshots and simulation boxes, it represents a valuable resource for future machine learning studies. The stellar circularity-based classification we developed here could serve as a baseline for benchmarking, or even as a training set for supervised learning models aimed at predicting galaxy morphology from other observable or derived properties. Such applications could help bridge the gap between simulation-based classifications and those used in observational surveys, facilitating domain adaptation in a cosmological context.

\section{CONCLUSION}\label{conclusion}

\indent In this preliminary study, we investigated whether stellar orbital circularities from the publicly available (c) \emph{Stellar Circularities, Angular Momenta, Axis Ratios} catalog in IllustrisTNG can be used as a basis for a broad morphological classification. After validating this approach against the more detailed kinematic decomposition in TNG50, we demonstrated that circularity-based disk and bulge fractions correlate well with physically motivated structural components, most notably, with the thin disk and the combined spherical subsystems (classical bulge and stellar halo), respectively.\\
\indent Using a data-informed threshold of $\mathrm{F_{disk}} = 0.25$, we classified galaxies in the TNG100 simulation into early-type, late-type, and irregular/complex systems. We showed that this threshold better captures the structural diversity of disk galaxies than the commonly adopted $\mathrm{F_{disk}} = 0.4$. In addition, we recovered a morphology–density relation that is broadly consistent with the observations, supporting the reliability of this simple classification scheme.\\
\indent Our results highlight the potential of circularity-based decomposition as a lightweight and scalable tool for large statistical studies of galaxy morphology in cosmological simulations. Although this method cannot capture fine morphological distinctions, its simplicity and availability across all snapshots and simulation volumes make it a valuable resource. It could be useful, especially for future studies involving machine learning, where it could serve as a baseline or training dataset.\\
\indent Future work may focus on refining this approach by combining circularity-based morphology with other global galaxy properties (such as gas content, star formation rate, and color), testing its robustness across redshift, and applying it to other simulation suites. A more detailed comparison with mock observations and observational morphology indicators will also be necessary to improve its applicability in bridging theory and observation.


\acknowledgements{The authors are grateful to the IllustrisTNG team for making their simulations publicly available. Python packages \texttt{matplotlib} \citep{Hunter2007}, \texttt{seaborn} \citep{Waskom2021}, \texttt{numpy} \citep{Harris2020}, and \texttt{pandas} \citep{McKinney2010} were used in parts of this analysis. K.B. acknowledges that this work was carried out as part of a summer internship made possible through the collaboration between the Faculty of Mathematics (University of Belgrade) and the Astronomical Observatory in Belgrade. A.M. acknowledges the support from the Ministry of Science, Technological Development and Innovation of the Republic of Serbia (MSTDIRS) through contract no. 451-03-136/2025-03/200002, made with the Astronomical Observatory (Belgrade, Serbia).}



\newcommand\eprint{in press }

\bibsep=0pt

\bibliographystyle{aa_url_saj}

{\small

\bibliography{sample_saj}
}

\begin{strip}

\end{strip}

\clearpage

{\ }

\clearpage

{\ }

\newpage

\begin{strip}

{\ }



\naslov{\MakeUppercase{Morfolo{\ss}ka klasifikacija galaksija: Da li su zvezdane cirkularnosti dovoljne?}}


\authors{Katarina Baucalo$^{1}$ and Ana Mitra{\v s}inovi{\' c}$^{2}$}

\vskip3mm


\address{$^1$Department of Astronomy, Faculty of Mathematics,
University of Belgrade\break Studentski trg 16, 11000 Belgrade,
Serbia}


\Email{kataarina981@gmail.com}

\address{$^2$Astronomical Observatory, Volgina 7, 11060 Belgrade 38, Serbia}

\Email{amitrasinovic@aob.rs}

\vskip3mm


\centerline{{\rrm UDK} \udc}


\vskip1mm

\centerline{\rit Uredjivaqki prilog}

\vskip.7cm

\baselineskip=3.8truemm

\begin{multicols}{2}

{
\rrm

Predstavljamo preliminarnu studiju u kojoj istra{\zz}ujemo da li se orbitalna cirkularnost zvezda u simuliranim galaksijama, dostupna kroz unapred izraqunate kataloge u okviru projekta} IllustrisTNG{\rrm , mo{\zz}e koristiti kao pokazatelj za grubu morfolo{\ss}ku klasifikaciju. Fokusiramo se na javno dostupan katalog "}Stellar Circularities, Angular Momenta, Axis Ratios{\rrm ", koji omogu{\cc}ava jednostavnu kinematiqku dekompoziciju zvezdanih komponenti na disk i sferoidne podsisteme. Validacijom ovog pristupa u odnosu na detaljniju dekompoziciju u pet komponenti iz} TNG50 {\rrm simulacije, potvrdjujemo da udeo diska zasnovan na cirkularnosti najbolje korelira sa tankim diskom, dok udeo ovala u {\ss}irem smislu predstavlja zbirni doprinos klasiqnih ovala i zvezdanih haloa. Ovu dekompoziciju zatim primenjujemo na galaksije u simulaciji} TNG100 {\rrm na crvenom pomaku $z = 0$ i identifikujemo prag $\mathrm{F_{disk}} = 0.25$, informisan podacima, kao kriterijum za razlikovanje ranih i kasnih morfolo{\ss}kih tipova. Ovaj prag, koji je ni{\zz}i od qesto kori{\ss}{\cc}ene vrednosti od 0.4, bolje obuhvata raznolikost diskolikih sistema i izbegava iskljuqivanje galaksija sa umereno izra{\zz}enim diskovima. Takodje identifikujemo nepravilne ili morfolo{\ss}ki kompleksne sisteme kao one sa malim ukupnim udelima diska i sferoidnih komponenti. Kori{\ss}{\cc}enjem ove klasifikacije, uspevamo da reprodukujemo morfologija-gustina relaciju koja je u saglasnosti sa posmatranjima: kasni tipovi galaksija preovlađuju u retkim sredinama (polju), dok su rani tipovi najzastupljeniji u jatima galaksija. Na{\ss}i rezultati pokazuju da sama zvezdana cirkularnost mo{\zz}e poslu{\zz}iti kao pristupaqan i raqunarski efikasan morfolo{\ss}ki pokazatelj. Takodje razmatramo potencijal ove klasifikacije da poslu{\zz}i kao osnova ili skup za treniranje u budu{\cc}im studijama morfolo{\ss}ke klasifikacije pomo{\cc}u metoda ma{\ss}inskog uqenja.

{\ }

}

\end{multicols}

\end{strip}


\end{document}